\documentclass{article}
\usepackage{fortschritte}
\usepackage{epsf}
\def\beq{\begin{equation}}                      
\def\eeq{\end{equation}}                       
\def\bea{\begin{eqnarray}}                             
\def\eea{\end{eqnarray}}

\def\cR{\mathcal R}
\def\lb{\bar{\lambda}}
\def\ub{\bar{u}}
\def\V{\vartheta}
\newcommand{\p}[4]{\Phi^{#1}_{#2}({#3})^{\mbox{\rm \scriptsize {#4}}}}
\newcommand{\lh}{\breve{\lambda}}
\newcommand{\uh}{\breve{u}}
\newcommand{\Fbeta}{\mbox{\boldmath $\beta$}}

\begin {document}   
\begin{center}
\hfill hep-th/0311056\\
\hfill MZ-TH/03-19\\
\hfill FSU-TPI-11/03
\end{center}              
\def\titleline{
%
%
%
%
Nonlocal Quantum Gravity             
and the \newtitleline                             
Size of the Universe\footnote{Talk given by M.R. at the 36th International Symposium Ahrenshoop, Berlin, August 26-30, 2003.} 
} 
%
%
\def\email_speaker{
{\tt 
%
%
speaker@engine.institute.country             
}}
\def\authors{
%
%
%
%
%
M. Reuter\1ad, F. Saueressig\2ad 
}
\def\addresses{
%
%
%
%
%
\1ad                                                   
Institute of Physics, University of Mainz \\          
Staudingerweg 7, D-55099 Mainz,  Germany \\            
\2ad                                        
Institute of Theoretical Physics, University of Jena \\            
Max-Wien-Platz 1, D-07743 Jena, Germany  \\           
%
}
\def\abstracttext{
%
%
Motivated by the conjecture that the cosmological constant problem is solved by strong quantum effects in the infrared we use the exact flow equation of Quantum Einstein Gravity to determine the renormalization group behavior of a class of nonlocal effective actions. They consist of the Einstein-Hilbert term and a general nonlinear function $F_k(V)$ of the Euclidean spacetime volume $V$. For the $V + V \ln V$-invariant the renormalization group running enormously suppresses the value of the renormalized curvature which results from Planck-size parameters specified at the Planck scale. One obtains very large, i.e., almost flat universes without finetuning the cosmological constant. A critical infrared fixed point is found where gravity is scale invariant. 
}
\large
\makefront
\section{Introduction}
Recently a lot of work on Quantum Einstein Gravity (QEG) went into constructing an appropriate exact renormalization group (RG) equation \cite{ERGE,Oliver_1}, finding approximate solutions to it \cite{Frank_1,Oliver_2,Dou,Frank_2,Souma}, and exploring their implications for black hole physics \cite{Alfio_1} and cosmology \cite{Alfio_2}. In particular, strong indications were found that QEG might be nonperturbatively renormalizable. If so, it has the status of a fundamental quantum theory of gravity, valid at arbitrarily short distances even. Here ``Einstein Gravity'' is supposed to mean a theory whose basic degrees of freedom are those of the metric $g_{\mu\nu}$. It is, however, {\it not} meant to imply that its bare action is of the traditional Einstein-Hilbert form, i.e., ``QEG'' is not an attempt at quantizing General Relativity. In fact, the RG approach tries to {\it compute} a bare action $\Gamma_\infty[g_{\mu\nu}]$ with the property that the quantum field theory based upon this highly non-generic action is nonperturbatively renormalizable \cite{wein}.

The basic tool used in these investigations is the effective average action and its exact ``Wilsonian'' RG equation \cite{wett}. The idea is to integrate out all fluctuation modes which have momenta larger than a certain infrared (IR) cutoff $k$ and to take account of those modes in an implicit way by the modified dynamics which they induce for the remaining fluctuations with momenta smaller than $k$. This ``renormalized'' dynamics is encoded in a scale dependent effective action $\Gamma_k$, whose dependence on the cutoff scale $k$ is governed by a functional differential equation referred to as the ``exact RG equation''. This equation gives rise to a flow on the space of all actions (``theory space''). The functional $\Gamma_k$ defines an effective field theory valid near the scale $k$; evaluated at {\it tree} level, it describes all loop effects due to the high momentum modes. 

The effective average action $\Gamma_k$, regarded as a function of $k$, interpolates between the ordinary effective action $\Gamma = \lim_{k \rightarrow 0} \Gamma_k$ and the bare (classical) action $S$ which is approached for $k \rightarrow \infty$. The construction of $\Gamma_k$ begins by adding a IR cutoff term $\Delta_kS$ to the classical action entering the standard Euclidean functional integral for the generating functional $W$ of the connected Green's functions. The new piece $\Delta_k S$ introduces a momentum dependent $\mbox{(mass)}^2$-term $\cR_k(p^2)$ for each mode of the quantum field with momentum $p$. For $p^2 \gg k^2$, the cutoff function $\cR_k(p^2)$ is assumed to vanish so that the high-momentum modes get integrated out unsuppressed. For $p^2 \ll k^2$, it behaves as $\cR_k(p^2) \propto k^2$; hence the small-momentum modes are suppressed in the path integral by a mass term $\propto k^2$. Up to a correction term known explicitly, the effective average action $\Gamma_k$ is the Legendre transform of the modified generating functional $W_k$.
From this definition one can derive the exact RG equation obeyed by $\Gamma_k$. In a slightly symbolic notation it is of the form
\beq\label{1.1}
k \, \partial_k \, \Gamma_k = \frac{1}{2} \, {\rm Tr} \left[ \left( \Gamma_k^{(2)} + \cR_k(-\Delta) \right)^{-1} \, k \, \partial_k \cR_k(- \Delta) \right] \, ,
\eeq              
where $\Gamma^{(2)}_k$ denotes the Hessian of $\Gamma_k$. The RHS of this equation is a kind of ``$\Fbeta$-functional'' which summarizes the $\Fbeta$-functions for infinitely many running couplings. Geometrically, it defines a vector field on theory space, the corresponding flow lines being the RG trajectories $k \mapsto \Gamma_k$.
In (\ref{1.1}) the c-number argument of $\cR_k$ is replaced with the operator $-\Delta$, typically a (generalized) covariant Laplacian. The discrimination of high-``momentum'' vs. low-``momentum'' modes is performed according to the spectrum of this operator, i.e., $p^2$ is an eigenvalue of $- \Delta$. 

\section{The Einstein-Hilbert Truncation}
\begin{figure}[t]

\epsfxsize=0.99 \textwidth
\centerline{\epsfbox{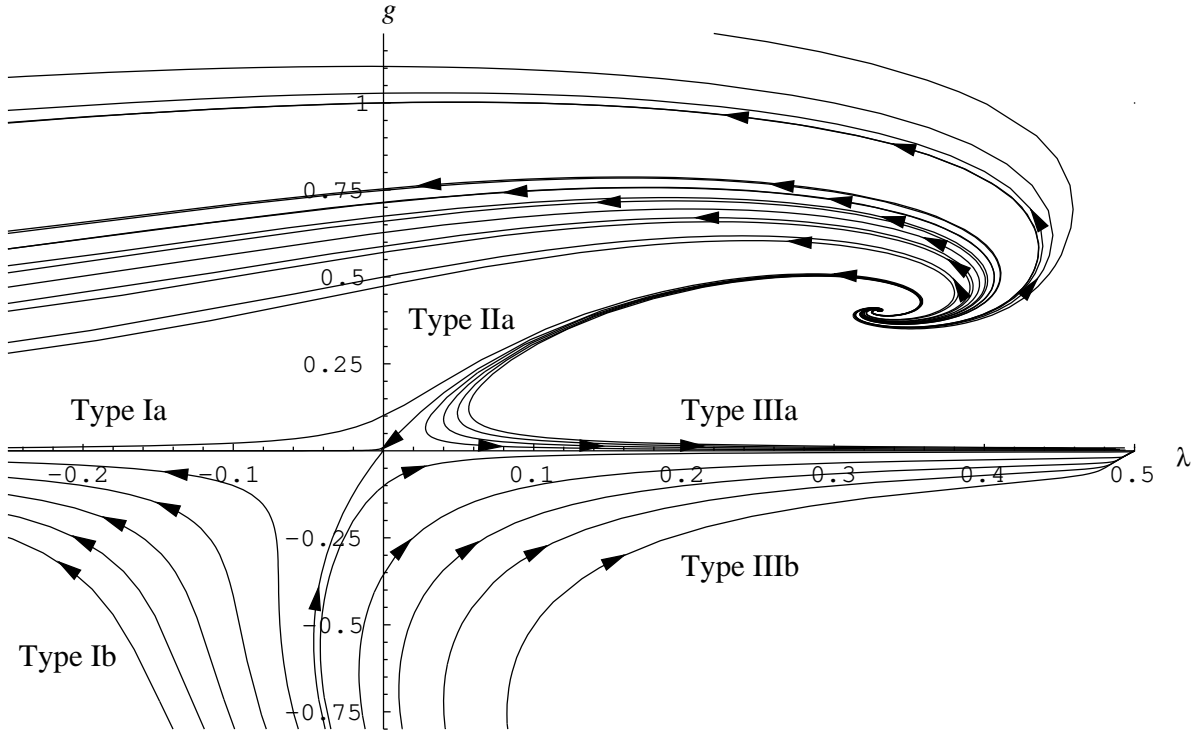}}
\parbox[c]{\textwidth}{\caption{\label{eins}{Part of coupling constant space of the Einstein-Hilbert truncation with its RG flow. The arrows point in the direction of decreasing values of $k$. The flow pattern is dominated by a non-Gaussian fixed point in the first quadrant and a trivial one at the origin.}}}
\end{figure}

Nonperturbative solutions to eq. (\ref{1.1}) can be obtained by ``truncating'' the theory space, i.e., by projecting the RG flow onto a ``small'', often finite-dimensional subspace. As for QEG, the first truncation studied \cite{ERGE} was the Einstein-Hilbert truncation defined by the ansatz
$
\Gamma_k \left[ g_{\mu \nu} \right] = \left( 16 \pi G_k \right)^{-1} \int d^4x \sqrt{g} \left\{ - R(g) + 2 \lb_k \right\} \, .
$
Inserting it into (\ref{1.1}) one obtains a system of two coupled differential equations governing the $k$-dependence of the running Newton constant $G_k$ and cosmological constant $\lb_k$. Reexpressing them in terms of the dimensionless couplings $g(k) \equiv k^2 G_k$ and $\lambda(k) \equiv \lb_k / k^2$ this system can be solved numerically \cite{Frank_1}. The resulting flow on the $g$-$\lambda$-plane is shown in Fig. \ref{eins}. 
It is dominated by two fixed points $(g_*, \lambda_* )$: a Gaussian fixed point (GFP) at $g_* = \lambda_* = 0$, and a non-Gaussian fixed point (NGFP) with $g_* > 0$ and $\lambda_* > 0$ \cite{Souma}. There are three classes of trajectories emanating from the NGFP: trajectories of type Ia and IIIa run towards negative and positive cosmological constants, respectively, and the single trajectory of type IIa (``separatrix'') hits the GFP for $k \rightarrow 0$. The short-distance properties of QEG are governed by the NGFP; for $k \rightarrow \infty$, in Fig. \ref{eins} all RG trajectories on the half-plane $g > 0$ run into this point. The conjectured nonperturbative renormalizability of QEG is due to the NGFP: if it is present in the full RG equations, it can be used to construct a microscopic quantum theory of gravity by taking the limit of infinite UV cutoff along one of the trajectories running into the NGFP, thus being sure that the theory does not develop uncontrolled singularities at high energies \cite{wein}. Loosely speaking, {\it QEG is defined to be the theory whose bare action $S$ equals the fixed point action $\lim_{k \rightarrow \infty} \Gamma_k[g_{\mu\nu}]$}. The presence of the NGFP has been confirmed within a more general truncation including a $(\mbox{curvature})^2$-term \cite{Oliver_2}. In the proximity of the NGFP, the $g$-$\lambda$-projection of the more general flow is well approximated by the Einstein-Hilbert truncation. The fixed point has a small $R^2$-admixture, though, so that the bare action is not the one of standard general relativity. Using a different approximation, further evidence for the NGFP was found in \cite{max}.

For trajectories of type Ia and IIa the RG equations of the Einstein-Hilbert truncation are mathematically well defined all the way down from $k = \infty$ to $k = 0$, those of type IIIa terminate at a finite $k = k_{\rm term} > 0$, however, since the $\Fbeta$-functions diverge when the dimensionless cosmological constant approaches $1/2$. In fact, the line $\lambda = 1/2$ is a boundary of the ``theory space'' of Fig. \ref{eins}. While the Einstein-Hilbert truncation is believed to be reliable near the NGFP, the aborting trajectories are a typical symptom (known from QCD, for instance) indicating that the truncation becomes insufficient in the IR.

\section{The $F(V)$-Truncations}
In Ref. \cite{Frank_2} we made a first attempt at including nonlocal invariants into the truncation. Since the ordinary effective action $\Gamma = \Gamma_{k = 0}$ is known to contain nonlocal invariants it is clear that such terms must be generated during the evolution if they are not already present in $\Gamma_\infty$. It is also clear that, in a cosmological application, curvature polynomials built from terms like $\int d^4x \sqrt{g} R^n$ are unimportant in the infrared of a ``large'' universe whose curvature is small, and nonlocal terms are crucial therefore \cite{cw}. For a variety of reasons it is very desirable to get a better understanding of the IR properties of QEG. For instance, in \cite{Frank_2} we speculated that the cosmological constant problem can perhaps be solved by strong nonperturbative renormalization effects in the IR, possibly similar to those occurring in Yang-Mills theory. The question is why the present size of the Universe (its present Hubble radius, say) is so much larger than the Planck length. In fact, the ``natural'' value of the cosmological constant is of order $m_{\rm Pl}^2$, and if the classical Einstein equation is valid, it predicts a ``size'' of order $\ell_{\rm Pl} \equiv \sqrt{G} \equiv m_{\rm Pl}^{-1}$.

Motivated by their technical simplicity and the fact that they play some role in wormhole physics \cite{tv} also, we considered truncations of the from \cite{Frank_2}
\beq\label{3.1}
\Gamma_k \left[ g \right] = - \, \frac{1}{16 \pi G} \int d^dx \sqrt{g} R + \frac{1}{8 \pi G} F_k(V[g]) \, .
\eeq
Here $V[g] \equiv \int d^dx \sqrt{g}$ is the volume of the $d$-dimensional Euclidean spacetime, $F_k$ is an arbitrary scale dependent function, and $G$ is the ordinary Newton constant whose evolution is neglected here. Since we restrict ourselves to $k \le m_{\rm Pl}$ this should be a reliable approximation.
To start with, we made the following more restrictive two-parameter ansatz for $F_k$ which was inspired by the work of Taylor and Veneziano \cite{tv}:
$
F_k(V) = \lb_k V + \frac{1}{2} \ub_k V \ln(V/V_0) \, .
$
In terms of the dimensionless parameters $\lh(y) \equiv \lb_k / m_{\rm Pl}^2$, $\uh(y) \equiv \ub_k / m_{\rm Pl}^2$ with the scale variable $ y \equiv k^2 / m_{\rm Pl}^2$ the corresponding RG equations read in $d = 4$:
\bea\label{3.3}
\nonumber
\frac{d \lh(y)}{d y} & = & \frac{y}{2 \pi} \left\{ - 5 \ln\left[ 1 - 2 \lh(y)/y - \uh(y)/y \right] + 2 \zeta(3) \right\} \, , \\
\frac{d \uh(y)}{d y} & = & \frac{5}{\pi} \, y \, \frac{\uh(y)}{y - 2 \lh(y) - \uh(y)} \, .
\eea
The classification of the RG trajectories implied by (\ref{3.3}) leads to similar classes as in the Einstein-Hilbert truncation. A new feature is the existence of trajectories reaching the IR ($k=0$) with a {\it positive} renormalized cosmological constant $\lim_{k \rightarrow 0} \lb_k$. In this respect it improves upon the Einstein-Hilbert truncation. There still exist aborting trajectories though.

A rather exciting feature of this nonlocal truncation becomes obvious when one looks at the effective equations of motion $\delta \Gamma_k / \delta g_{\mu\nu} = 0$. Ignoring the $k$-dependence for a moment, it reads $R_{\mu\nu} - \frac{1}{2} \, g_{\mu\nu} R = - \lambda_{\rm eff}(V) g_{\mu \nu}$ where the effective cosmological constant $\lambda_{\rm eff}(V) \equiv \lb + \frac{1}{2} \ub + \frac{1}{2} \ub \ln(V/V_0)$ is metric dependent. The simplest solutions are spheres $S^4$ whose radius $r = r(\lb, \ub)$ is given by the transcendental equation $\ub r^2 + \ub r^2 \ln( \sigma r^4 / V_0) + 2 \lb r^2 - 6 = 0$ with $\sigma \equiv 8 \pi^2/3$. Assuming $\ub > 0$, one finds the following relation between the cosmological constant proper, $\lb$, and the effective cosmological constant. If $\lb > 0$, $\lambda_{\rm eff} \approx \lb$, but if $\lb < 0$, $\lambda_{\rm eff} \propto \exp\left(- |\lb|/\ub \right)$, so that $\lambda_{\rm eff}$ can be much smaller that $|\lb|$. Since, by Einstein's equation, $r = \left(3 / \lambda_{\rm eff} \right)^{1/2}$, this implies that the radius of the $S^4$ can be large compared to the scale $1/|\lb|$.

In \cite{Frank_2} we investigated the ``RG improvement'' of this mechanism. We considered $\lb \equiv \lb_k$, $\ub \equiv \ub_k$ running quantities whose $k$-dependence is given by the system (\ref{3.3}). In this manner the radius of the $S^4$ becomes $k$-dependent as well: $ r = r(\lb_k, \ub_k) \equiv r(k)$. For a Euclidean universe of radius $r$ the relevant effective action is $\Gamma_k$ at $k \approx 1/r$. As a consequence, when we explore the possibility for large universes ($r \rightarrow \infty$) we must use $\Gamma_{k = 0}$, i.e., the {\it renormalized} couplings $\lb_0$, $\ub_0$ rather than the {\it bare} ones, $\lb_{\hat{k}}$ and $\ub_{\hat{k}}$. Here $\hat{k}$ is the initial scale from which we evolve downward; usually we chose $\hat{k} = m_{\rm Pl}$. We solved the RG equations (\ref{3.3}) for a wide range of initial values of $\lh$ and $\uh$, all approximately of order unity, specified at the initial point $y = 1$. Hence the initial values of $\lb_k$ and $\ub_k$, fixed at $k = m_{\rm Pl}$, are of Planckian size, as it would be considered natural. For a given trajectory $k \mapsto (\lb_k, \ub_k)$ of the modified type Ia, with $\lb_0 < 0$, we then solved the transcendental equation for the radius $r(k)$. Comparing the ``bare'' radius $r(k = m_{\rm Pl})$ to the ``renormalized'' one, $r(k = 0)$, it turns out that the inclusion of the RG evolution leads to a tremendous ``inflation'' of the universe: $r(k = 0) \gg r(k = m_{\rm Pl})$. Starting from natural initial conditions at the Planck scale, it is very easy to obtain enhancement factors $r(k = 0)/r(k = m_{\rm Pl})$ of the order of $10^{100}$. This ``RG improved Taylor-Veneziano mechanism'' provides a very promising approach to explain the smallness of the (positive) cosmological constant we observe today. The $V \ln V$-example shows that, at least in principle, {\it a tiny nonlocal coupling in the effective action can effectively solve the cosmological constant problem even though the cosmological constant proper is not small}. (See ref. \cite{Frank_2} for further details.)

\section{The Universe as a ``Critical Phenomenon''}
Allowing for an arbitrary evolving function $F_k(V)$ is a more ambitious task. Projecting the RG flow on the infinite dimensional truncated theory space spanned by the ansatz (\ref{3.1}) one obtains a partial differential equation for $F_k(V)$. In terms of the dimensionless variables $f_k(\V) \equiv k^{d-2} F_k(k^{-d} \V)$, $\V \equiv k^d V$, and $g_k = k^{d-2} G$, it is of the from $k \, \partial_k \, g_k = \Fbeta_g(g_k, f_k)$, $k \, \partial_k \, f_k(\V) = \Fbeta_f(g_k, f_k)$. Since we neglect the running of $G$, its $\Fbeta$-function is the canonical one: $\Fbeta_g = (d-2) g_k$. For $\Fbeta_f$ one finds \cite{Frank_2}
\bea\label{4.1}
\Fbeta_f(g, f) &=& - \left\{ (2-d) \, f_k(\vartheta) + d \, \vartheta \, f_k^{\prime}(\vartheta) \right\} \\
\nonumber && + 8 \, \pi \, g \, \bigg\{ 
(4 \pi )^{-d/2} \, \vartheta \, \left( \frac{d \, (d+1)}{2} \, \p{1}{d/2}{-2 \, f_k^{\prime}(\vartheta)}{} - 2 \, d \, \p{1}{d/2}{0}{} \right) \hspace{1cm} \\ 
\nonumber && \qquad \qquad - \, \frac{1}{1 - 2 \, f_k^{\prime}(\vartheta)} + \frac{1}{1 - 2 \, f_k^{\prime}(\vartheta) - \frac{2d}{d-2} \, f_k^{\prime \prime}(\vartheta) \, \vartheta } 
\bigg\}
\eea
where the $\Phi$'s denote the threshold functions introduced in \cite{ERGE}. This flow possesses a single fixed point: the modified Gaussian fixed point (MGFP) with $g_* = 0$ and the ``fixed function'' $f_*(\V) = c \, \V^{(d-2)/d}$. Here c is an arbitrary dimensionless constant which actually parametrizes a one-parameter family of fixed points. The associated $\Gamma[g]$ is $k$-independent; in $d=4$, 
\beq\label{4.2}
\Gamma_*[g] = \frac{1}{16 \pi G} \, \int d^4x \, \sqrt{g} (-R) + \frac{c}{8 \pi G} \, \left( \int d^4x \sqrt{g} \right)^{1/2} \, .
\eeq
The action $\Gamma_*[g]$ has various properties which are quite remarkable: (i) $\Gamma_*$ has vanishing cosmological constant. (ii) The action $\Gamma_*$ is scale invariant in the sense that the associated equation of motion $R_{\mu\nu} - \frac{1}{2} g_{\mu\nu} R = - c g_{\mu\nu} / \sqrt{4 V[g]}$ does not contain any dimensionful coupling constant. Hence this equation cannot fix the ``size'' of the universe. Inserting a $S^4$-ansatz, say, the radius drops out and one finds that spheres of {\it any} radius are solutions provided $c = 12 \pi  \sqrt{2/3}$. (iii) A detailed stability analysis reveals that, on the $\{F(\cdot)\}$-theory space, the MGFP is marginally stable for $k \rightarrow 0$, i.e., it has no IR-repulsive directions. (This is to be contrasted with the ordinary GFP which is strongly repulsive in $\lambda$-direction.) As a result, trajectories which are to run into the MGFP for $k \rightarrow 0$ do not require any finetuning of their initial conditions.

The MGFP is a simple example of a critical fixed point at which the system has no distinguished length scale. If the RG trajectory followed by the universe is attracted into the MGFP for $k \rightarrow 0$, its actual size is completely unrelated to the value of the bare cosmological constant $\lb_{\hat{k}}$.

\section{Conclusion}
The $F(V)$-truncation is certainly not realistic yet, but it demonstrates that the smallness of the observed, i.e., renormalized, cosmological constant can indeed be the result of a nonperturbative gravitational RG evolution in the infrared, and that the actual large-distance properties of the universe in general have little or nothing to do with the naive, i.e., bare, values of the gravitational parameters.


\end{document}